\documentclass[twocolumn,showpacs,preprintnumbers,amsmath,amssymb,prd,superscriptaddress]{revtex4-1}

\usepackage{amsmath}
\usepackage{graphicx}
\usepackage{dcolumn}
\usepackage{bm}
\usepackage{multirow}
\usepackage{array}
\usepackage{tabularx}
\usepackage{xcolor}
\usepackage{lipsum}

\usepackage[mathlines]{lineno}
\graphicspath{{image/}}

\begin{document}


\title{Transverse spin transfer to $\Lambda$ and $\bar{\Lambda}$ hyperons in polarized proton-proton collisions \\at $\sqrt{s}=200\,\mathrm{GeV}$}

\affiliation{Abilene Christian University, Abilene, Texas   79699}
\affiliation{AGH University of Science and Technology, FPACS, Cracow 30-059, Poland}
\affiliation{Alikhanov Institute for Theoretical and Experimental Physics, Moscow 117218, Russia}
\affiliation{Argonne National Laboratory, Argonne, Illinois 60439}
\affiliation{Brookhaven National Laboratory, Upton, New York 11973}
\affiliation{University of California, Berkeley, California 94720}
\affiliation{University of California, Davis, California 95616}
\affiliation{University of California, Los Angeles, California 90095}
\affiliation{University of California, Riverside, California 92521}
\affiliation{Central China Normal University, Wuhan, Hubei 430079 }
\affiliation{University of Illinois at Chicago, Chicago, Illinois 60607}
\affiliation{Creighton University, Omaha, Nebraska 68178}
\affiliation{Czech Technical University in Prague, FNSPE, Prague 115 19, Czech Republic}
\affiliation{Technische Universit\"at Darmstadt, Darmstadt 64289, Germany}
\affiliation{E\"otv\"os Lor\'and University, Budapest, Hungary H-1117}
\affiliation{Frankfurt Institute for Advanced Studies FIAS, Frankfurt 60438, Germany}
\affiliation{Fudan University, Shanghai, 200433 }
\affiliation{University of Heidelberg, Heidelberg 69120, Germany }
\affiliation{University of Houston, Houston, Texas 77204}
\affiliation{Indiana University, Bloomington, Indiana 47408}
\affiliation{Institute of Modern Physics, Chinese Academy of Sciences, Lanzhou, Gansu 730000 }
\affiliation{Institute of Physics, Bhubaneswar 751005, India}
\affiliation{University of Jammu, Jammu 180001, India}
\affiliation{Joint Institute for Nuclear Research, Dubna 141 980, Russia}
\affiliation{Kent State University, Kent, Ohio 44242}
\affiliation{University of Kentucky, Lexington, Kentucky 40506-0055}
\affiliation{Lawrence Berkeley National Laboratory, Berkeley, California 94720}
\affiliation{Lehigh University, Bethlehem, Pennsylvania 18015}
\affiliation{Max-Planck-Institut f\"ur Physik, Munich 80805, Germany}
\affiliation{Michigan State University, East Lansing, Michigan 48824}
\affiliation{National Research Nuclear University MEPhI, Moscow 115409, Russia}
\affiliation{National Institute of Science Education and Research, HBNI, Jatni 752050, India}
\affiliation{National Cheng Kung University, Tainan 70101 }
\affiliation{Nuclear Physics Institute AS CR, Prague 250 68, Czech Republic}
\affiliation{Ohio State University, Columbus, Ohio 43210}
\affiliation{Institute of Nuclear Physics PAN, Cracow 31-342, Poland}
\affiliation{Panjab University, Chandigarh 160014, India}
\affiliation{Pennsylvania State University, University Park, Pennsylvania 16802}
\affiliation{Institute of High Energy Physics, Protvino 142281, Russia}
\affiliation{Purdue University, West Lafayette, Indiana 47907}
\affiliation{Pusan National University, Pusan 46241, Korea}
\affiliation{Rice University, Houston, Texas 77251}
\affiliation{Rutgers University, Piscataway, New Jersey 08854}
\affiliation{Universidade de S\~ao Paulo, S\~ao Paulo, Brazil 05314-970}
\affiliation{University of Science and Technology of China, Hefei, Anhui 230026}
\affiliation{Shandong University, Qingdao, Shandong 266237}
\affiliation{Shanghai Institute of Applied Physics, Chinese Academy of Sciences, Shanghai 201800}
\affiliation{Southern Connecticut State University, New Haven, Connecticut 06515}
\affiliation{State University of New York, Stony Brook, New York 11794}
\affiliation{Temple University, Philadelphia, Pennsylvania 19122}
\affiliation{Texas A\&M University, College Station, Texas 77843}
\affiliation{University of Texas, Austin, Texas 78712}
\affiliation{Tsinghua University, Beijing 100084}
\affiliation{University of Tsukuba, Tsukuba, Ibaraki 305-8571, Japan}
\affiliation{United States Naval Academy, Annapolis, Maryland 21402}
\affiliation{Valparaiso University, Valparaiso, Indiana 46383}
\affiliation{Variable Energy Cyclotron Centre, Kolkata 700064, India}
\affiliation{Warsaw University of Technology, Warsaw 00-661, Poland}
\affiliation{Wayne State University, Detroit, Michigan 48201}
\affiliation{Yale University, New Haven, Connecticut 06520}

\author{J.~Adam}\affiliation{Creighton University, Omaha, Nebraska 68178}
\author{L.~Adamczyk}\affiliation{AGH University of Science and Technology, FPACS, Cracow 30-059, Poland}
\author{J.~R.~Adams}\affiliation{Ohio State University, Columbus, Ohio 43210}
\author{J.~K.~Adkins}\affiliation{University of Kentucky, Lexington, Kentucky 40506-0055}
\author{G.~Agakishiev}\affiliation{Joint Institute for Nuclear Research, Dubna 141 980, Russia}
\author{M.~M.~Aggarwal}\affiliation{Panjab University, Chandigarh 160014, India}
\author{Z.~Ahammed}\affiliation{Variable Energy Cyclotron Centre, Kolkata 700064, India}
\author{I.~Alekseev}\affiliation{Alikhanov Institute for Theoretical and Experimental Physics, Moscow 117218, Russia}\affiliation{National Research Nuclear University MEPhI, Moscow 115409, Russia}
\author{D.~M.~Anderson}\affiliation{Texas A\&M University, College Station, Texas 77843}
\author{R.~Aoyama}\affiliation{University of Tsukuba, Tsukuba, Ibaraki 305-8571, Japan}
% ...

\collaboration{STAR Collaboration}\noaffiliation

\date{\today}

\begin{abstract}
The transverse spin transfer from polarized protons to $\Lambda$ and $\bar{\Lambda}$ hyperons is expected to provide sensitivity to the transversity distribution of the nucleon and to the transversely polarized fragmentation functions. 
We report the first measurement of the transverse spin transfer to $\Lambda$ and $\bar{\Lambda}$ along the polarization direction of the fragmenting quark, $D_\mathrm{TT}$, in transversely polarized proton-proton collisions at $\sqrt{s}=200\,\mathrm{GeV}$ with the STAR detector at RHIC. 
The data correspond to an integrated luminosity of $18\,\mathrm{pb}^{-1}$ and cover the pseudorapidity range $\left|\eta\right| < 1.2$ and transverse momenta $p_{\mathrm{T}}$ up to $8\,\mathrm{GeV}/c$. 
The dependence on $p_\mathrm{T}$ and $\eta$ are presented.
The $D_\mathrm{TT}$ results are found to be comparable with a model prediction, and are also consistent with zero within uncertainties.
\end{abstract}

\maketitle

The polarizations of $\Lambda$ and $\bar\Lambda$ hyperons have been studied extensively in various aspects of spin effects in high energy reactions due to the self spin-analyzing parity violating decay\,\cite{Lee:1957qs}.
Many of them have been found to be quite surprising, for example, the induced transverse polarization with respect to the production plane in unpolarized hadron-hadron reactions\,\cite{Bunce:1976yb}, and the global polarization with respect to the reaction plane observed recently in heavy ion collisions\,\cite{STAR:2017ckg}. 
On the other hand, $\Lambda$ polarization transferred from polarized lepton or hadron beams (usually referred to as ``spin transfer") in different reactions provides a natural connection to polarized fragmentation functions and the polarized parton densities of the nucleon. 
A number of measurements have been made in polarized lepton-nucleon deep inelastic scattering (DIS), for example, E665\,\cite{Adams:1999px}, HERMES\,\cite{Airapetian:2006ee} and COMPASS\,\cite{Alekseev:2009ab}, and in polarized hadron-hadron collisions in the STAR experiment at RHIC\,\cite{09DLL:2009xg,Adam:2018kzl}. 
In particular, among the polarized parton distributions, the transversity distribution remains much less known than the helicity distribution due to its chiral-odd nature\,\cite{Barone:2001sp,Perdekamp:2015vwa}. 
The transversity distribution describes the probability of finding a transversely polarized quark in a transversely polarized proton.
Interestingly, the transverse spin transfer to $\Lambda$ in lepton-hadron and/or hadron-hadron collisions provides a natural connection to transversity through polarized fragmentation functions\,\cite{Artru:1990wq,Jaffe:1996wp,deFlorian:1998am,Ma:2001rm,Xu:2004es,polarizationLA:2005ru}. 
Previously, sizable spin transfer along the normal direction of the $\Lambda$ production plane, $D_{\mathrm{NN}}$, was observed at large $x_{\mathrm{F}}$ with fixed target proton-proton collisions by the E704 Collaboration at Fermilab\,\cite{Bravar:1997fb} and also in exclusive production in proton-proton reactions at low energy\,\cite{Balestra:1999br}. 

We report the first measurement of transverse spin transfer to $\Lambda$ and $\bar\Lambda$ hyperons in transversely polarized proton-proton collisions at $\sqrt s=200$\,GeV with the STAR experiment at RHIC at Brookhaven National Laboratory. 
The transverse spin transfer, $D_\mathrm{TT}$, to the $\Lambda$ in proton-proton collisions is defined as:
\begin{equation}
  \centering
  D_\mathrm{TT}
  \equiv
  \frac{d\sigma^{{(p^{\uparrow}p \rightarrow \Lambda^{\uparrow}X)}}-d\sigma^{{(p^{\uparrow}p \rightarrow \Lambda^{\downarrow}X)}}} {d\sigma^{{(p^{\uparrow}p \rightarrow \Lambda^{\uparrow}X)}}+d\sigma^{{(p^{\uparrow}p \rightarrow \Lambda^{\downarrow}X)}}}
  =
  \frac{d\delta\sigma^{\Lambda}}{d\sigma^{\Lambda}},
  \label{eq:dttDef}
\end{equation}
where $\uparrow$\,($\downarrow$) denotes the positive\,(negative) transverse polarization direction of the particles and $\delta\sigma^{\Lambda}$ is the transversely polarized cross section.
Within the factorization framework,  $\delta\sigma^{\Lambda}$ can be factorized into the convolution of parton transversity, polarized partonic cross section and the polarized fragmentation function\,\cite{deFlorian:1998am}.
As the $s(\bar{s})$ quark plays a dominant role in $\Lambda\,(\bar{\Lambda})$ hyperon's spin content, the measurements of  $D_{\mathrm{TT}}$ for $\Lambda\,(\bar{\Lambda})$ hyperon provide a natural connection to the transversity distribution of strange and anti-strange quarks\, \cite{deFlorian:1998am,Xu:2004es,polarizationLA:2005ru}.
Experimentally, the transverse spin transfer, $D_{\mathrm{TT}}$, can be measured along the transverse polarization direction of the outgoing quark after the hard scattering. 

The polarization of $\Lambda\,(\bar{\Lambda})$ hyperons, $P_{\Lambda\,(\bar{\Lambda})}$, can be measured from the angular distribution of the final state particles via their weak decay channel $\Lambda\rightarrow p\pi^{-}\,(\bar{\Lambda}\rightarrow \bar{p}\pi^{+})$,
\begin{equation}
  \frac
  {dN}{\,d\cos{\theta^{*}}\,}
  \propto
  A
  \left(1+\alpha_{\Lambda\,(\bar{\Lambda})}P_{\Lambda\,(\bar{\Lambda})}\cos{\theta^{*}}\right),
  \label{eq:distFinalState}
\end{equation}
where $A$ is the detector acceptance varying with $\theta^{*}$ as well as other observables, $\alpha_{\Lambda\,(\bar{\Lambda})}$ is the weak decay parameter, and $\theta^{*}$ is the angle between the $\Lambda\,(\bar{\Lambda})$ polarization direction and the (anti-)\,proton momentum in the $\Lambda\,(\bar{\Lambda})$ rest frame. 
For the $D_{\mathrm{TT}}$ measurements in this analysis, the transverse polarization direction of the outgoing fragmenting parton is used to obtain $\theta^{*}$.
Since there is a rotation along the normal direction of the scattering plane between the transverse polarization directions of incoming and outgoing quarks\,\cite{JetPolar:1993kq}, the direction of the momentum of the outgoing parton is required and the reconstructed jet axis adjacent to the $\Lambda\,(\bar{\Lambda})$ is used as the substitute for the direction of the outgoing fragmenting quark.

The data were collected at RHIC with the STAR experiment in the year 2012.
An integrated luminosity of $18\,\mathrm{pb}^{-1}$ was sampled with transverse proton beam polarization.
The proton polarization was measured for each beam and per fill using Coulomb-Nuclear Interference\,(CNI) proton-Carbon polarimeters\,\cite{CNIpolarmeter:2006dd}, which were calibrated using a polarized atomic hydrogen gas-jet target\,\cite{pCpolar:2004up}.
The average transverse polarizations for the two beams were $58\%$ and $64\%$ for the analyzed data.
Polarization up and down bunch patterns were changed between beam fills to minimize systematic uncertainties. 

The primary detector sub-system used in this analysis is the Time Projection Chamber (TPC)\,\cite{TPC:2003ur}, which provides tracking for charged
particles in the 0.5 T magnetic field in the pseudo-rapidity range of $\left|\eta\right|<1.2$ with full azimuthal coverage.
The measurement of specific energy loss, $\mathrm{d}E/\mathrm{d}x$, in the TPC gas provided information for particle identification.
The Barrel Electromagnetic Calorimeter\,(BEMC)\,\cite{BEMC:2002zx} and Endcap Electromagnetic Calorimeter\,(EEMC)\,\cite{EEMC:2002zy} were used in generating the primary jet trigger information at STAR.
The total BEMC coverage was $\left|\eta\right|<1$ with full azimuth in 2012, and the EEMC extends the pseudo-rapidity coverage up to $\eta\sim2$.
The data samples used in this analysis were recorded with jet-patch\,(JP) trigger conditions which required a transverse energy deposit $E_{\mathrm{T}}$ in BEMC or EEMC patches (each covering a range of $\Delta\eta\times\Delta\phi = 1\times1$ in  pseudo-rapidity and azimuthal angle) exceeding certain thresholds.
The thresholds were $E_{\mathrm{T}}\sim3.5\,\mathrm{GeV}$ for JP0, or $E_{\mathrm{T}}\sim5.4\,\mathrm{GeV}$ for JP1, or $E_{\mathrm{T}}\sim7.3\,\mathrm{GeV}$ for JP2, or two adjacent jet patches (AJP) with each exceeding the threshold of $E_{\mathrm{T}}\sim3.5\,\mathrm{GeV}$ for the AJP trigger.

The $\Lambda$ and $\bar{\Lambda}$ candidates were reconstructed from their dominant weak decay channels, $\Lambda\rightarrow p\pi^{-}$ and $\bar{\Lambda}\rightarrow \bar{p}\pi^{+}$, with a branching ratio of $63.9\%$\,\cite{Tanabashi:2018pdg}.
The location of the primary vertex (PV) was required to be within 60\,cm of the center of the TPC along the beam axis to ensure uniform acceptance. 
The selection procedure of $\Lambda$ and $\bar{\Lambda}$ candidates in this analysis is based on the topology of the weak decay using a method that is very similar to the one used in the previous longitudinal spin transfer measurement reported in Ref.\,\cite{09DLL:2009xg}.
(Anti-)\,proton and pion tracks were first identified based on $\mathrm{d}E/\mathrm{d}x$ in the TPC.
They were paired to form a $\Lambda\,(\bar{\Lambda})$ candidate, and topological selections were used to further reduce the background.
The selection cuts were tuned in each hyperon $p_{\mathrm{T}}$ bin to keep the residual background fraction below 10\%; these cuts are summarized in Table~\ref{tab:NLA_J1}.
\begin{figure}[tbp]
  \centering
  \includegraphics[width=0.48\textwidth]{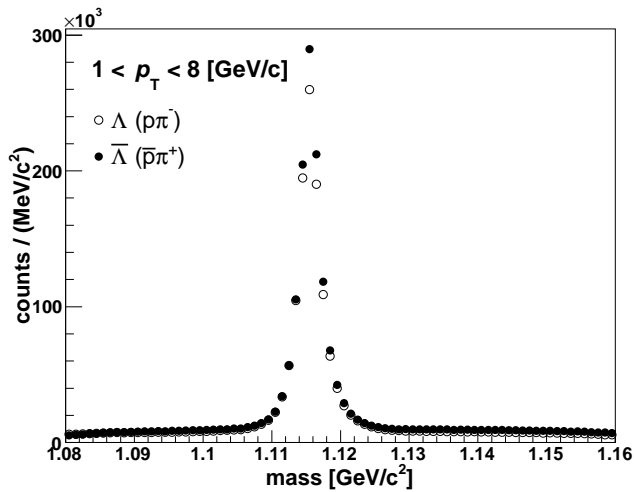}
  \caption{
  The invariant mass distribution for $\Lambda$ (open circles) and $\bar{\Lambda}$ (filled circles) candidates with $1<p_{\mathrm{T}}<8\,\mathrm{GeV}/c$ after selection in this analysis.
  }
  \label{fig:im}
\end{figure}

As mentioned earlier, reconstructed jets were employed to obtain the momentum direction of the fragmenting quark and thus the transverse polarization direction of hyperons for the $D_{\mathrm{TT}}$ measurement.
In this analysis, jets were reconstructed using the anti-$k_{\mathrm{T}}$ algorithm\,\cite{antiKt:2008gp} with a resolution parameter $R=0.6$ and jet $p_\mathrm{T}>5\,\mathrm{GeV}/c$.
In STAR, the TPC tracks and tower energies from the BEMC and EEMC are used for the jet reconstruction\,\cite{Adamczyk:2014ozi,Adamczyk:2016okk}. 
Then the association between $\Lambda\,(\bar{\Lambda})$ candidates and the adjacent reconstructed jet was made by constraining the radius,
$\Delta R=\sqrt{(\Delta\eta)^{2}+(\Delta\phi)^{2}}\,$,
between the $\Lambda\,(\bar{\Lambda})$ momentum direction and the jet axis in $\eta-\phi$ space.
The $\Lambda\,(\bar{\Lambda})$ candidates that have a jet with $\Delta R<0.6$ were used in the following $D_{\mathrm{TT}}$ determination.
The fraction of $\Lambda\,(\bar{\Lambda})$ candidates that have a matching jet increases from about 30\% to 90\% from the lowest hyperon $p_\mathrm{T}$ bin to the highest.
\begin{figure}[!tb]
  \centering
  \includegraphics[width=0.48\textwidth]{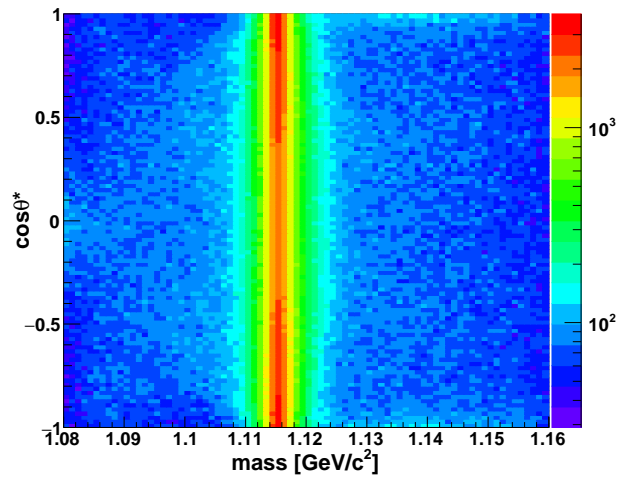}
  \caption{
  The invariant mass distribution versus $\cos{\theta^{*}}$ for $\Lambda$ candidates with $1<p_{\mathrm{T}}<8\,\mathrm{GeV}/c$ in this analysis as an example.
  }
  \label{fig:imCos}
\end{figure}

After topological selections and jet correlation, the invariant mass distributions for the $\Lambda$ and $\bar{\Lambda}$ candidates with $1<p_{\mathrm{T}}<8\,\mathrm{GeV}/c$ and $|\eta|<1.2$ are shown in Fig.\,\ref{fig:im}.
The values of the peak in the $\Lambda$ and $\bar{\Lambda}$ mass distributions are in agreement with the PDG mass value $m_{\Lambda\,(\bar{\Lambda})}=1.11568\,\mathrm{GeV}/c^{2}$\,\cite{Tanabashi:2018pdg}.
The 2-dimensional distribution of invariant mass versus $\cos{\theta^{*}}$ as defined in Eq.\,\ref{eq:distFinalState} is shown in Fig.\,\ref{fig:imCos} for $\Lambda$ candidates. 
A similar distribution was also obtained for $\bar{\Lambda}$ candidates.
The bin counting method was used to obtain the raw yields of $\Lambda$ and $\bar{\Lambda}$ candidates.
The signal mass windows were chosen to be about $3\sigma$ of the mass width.
These range from $1.111\sim1.119\,\mathrm{GeV}/c^{2}$ in the lowest $p_\mathrm{T}$ bin to $1.104\sim1.128\,\mathrm{GeV}/c^{2}$ in the highest $p_\mathrm{T}$ bin. 
About $1.02\times10^{6}$ $\Lambda$ and $1.17\times10^{6}$ $\bar{\Lambda}$ candidates in the selected signal mass windows were kept as the signal.
The residual background fractions were estimated by the side-band method and found to be $7\%\sim10\%$.
The yields, background fractions and signal mass windows are listed in Table~\ref{tab:NLA_J1}.
The $\bar{\Lambda}$ signal is larger than that of the $\Lambda$ because of the effects of anti-proton annihilation in the BEMC and EEMC, which provides additional energy to the JP trigger conditions for the $\bar{\Lambda}$.
\begin{table*}[hbpt]\scriptsize
  \renewcommand\arraystretch{1.5}
  \caption{Summary of selection cuts and the $\Lambda$ and $\bar{\Lambda}$ candidate counts and the residual background fractions in each $p_\mathrm{T}$ bin. Here ``DCA" denotes distance of closest approach, and $N(\sigma)$ quantitatively measures the distance of a particle track to a certain particle band in $\mathrm{d}E/\mathrm{d}x$ vs. rigidity space\,\cite{Abelev:2006cs}. $\vec{l}$ is representative of the vector from the PV to the $\Lambda$ decay point and $\vec{p}$ is the reconstructed momentum of $\Lambda$.}
  \resizebox{\textwidth}{!}{
  \begin{tabular}{c|cccccc}
    \hline\hline
    $p_{\mathrm{T}}\,[\mathrm{GeV/c}]$ &(1,\,2) &(2,\,3) &(3,\,4) &(4,\,5) &(5,\,6) &(6,\,8) \\
    \hline
    N(hits) of daughter tracks                           &\textgreater\,14 &\textgreater\,14 &\textgreater\,14 &\textgreater\,14 &\textgreater\,14 &\textgreater\,14  \\
    \hline
    N($\sigma$) $d\mathrm{E}/d\mathrm{x}$ for daughters  &\textless\,3  &\textless\,3  &\textless\,3  &\textless\,3  &\textless\,3  &\textless\,3   \\
    \hline
    DCA of daughter tracks\,[cm]                            &\textless\,0.80 &\textless\,0.70 &\textless\,0.60 &\textless\,0.50 &\textless\,0.45 &\textless\,0.45 \\
    \hline
    DCA of $\Lambda$\,($\bar{\Lambda}$) to PV\,[cm]                                &\textless\,1.0  &\textless\,1.0  &\textless\,1.0  &\textless\,1.0  &\textless\,1.0  &\textless\,1.0  \\
    \hline
    $\cos{(\vec{l}\cdot\vec{p})}$                            &\textgreater\,0.995 &\textgreater\,0.995 &\textgreater\,0.995 &\textgreater\,0.995 &\textgreater\,0.995 &\textgreater\,0.995 \\
    \hline
    Decay Length\,[cm]                                   &\textgreater\,3.5  &\textgreater\,4.0  &\textgreater\,4.0  &\textgreater\,4.5  &\textgreater\,5.0  &\textgreater\,5.0  \\
    \hline
    DCA of $p\,(\bar{p})$ to PV\,[cm]                    &\textgreater\,0.25 &\textgreater\,0.20 &\textgreater\,0.10 &\textgreater\,0.05 &\textgreater\,0.05 &\textgreater\,0.05 \\
    \hline
    DCA of $\pi^{-}(\pi^{+})$ to PV\,[cm]              &\textgreater\,0.60 &\textgreater\,0.55 &\textgreater\,0.50 &\textgreater\,0.50 &\textgreater\,0.50 &\textgreater\,0.50 \\
    \hline
    \hline
    $\Lambda\,(\bar{\Lambda})$ counts                    &469681\,(502226) &318358\,(368042) &181550\,(193221) &77866\,(71833) &32441\,(25571) &20256\,(13486) \\
    \hline
    $\Lambda\,(\bar{\Lambda})$ bkg. frac.                &0.065\,(0.074) &0.079\,(0.077) &0.071\,(0.072) &0.065\,(0.066) &0.070\,(0.075) &0.084\,(0.102) \\
    \hline
    Mass window(signal)\,[$\mathrm{GeV}/c^{2}$]                &(\,1.111,\,1.119\,) &(\,1.111,\,1.121\,) &(\,1.109,\,1.123\,) &(\,1.108,\,1.124\,) &(\,1.106,\,1.126\,) &(\,1.104,\,1.128\,) \\
    \hline
    Side-band (left)\,[$\mathrm{GeV}/c^{2}$]                &(\,1.090,\,1.100\,) &(\,1.090,\,1.100\,) &(\,1.088,\,1.098\,) &(\,1.087,\,1.097\,) &(\,1.085,\,1.095\,) &(\,1.083,\,1.093\,) \\
    \hline
    Side-band (right)\,[$\mathrm{GeV}/c^{2}$]                &(\,1.130,\,1.140\,) &(\,1.132,\,1.142\,) &(\,1.134,\,1.144\,) &(\,1.135,\,1.145\,) &(\,1.137,\,1.147\,) &(\,1.139,\,1.149\,) \\
    \hline\hline
  \end{tabular}
  }
  \label{tab:NLA_J1}
\end{table*}

The observed $\cos{\theta^{*}}$ spectra are affected by detector efficiency and acceptance as seen in Eq.\,\ref{eq:distFinalState}.
To minimize the systematics associated with acceptance and relative luminosity, $D_{\mathrm{TT}}$ has been extracted from the asymmetry in small $\cos{\theta^{*}}$ intervals using 
\begin{widetext}
\begin{equation}
  D_{\mathrm{TT}} = 
  \frac{1}{\,\alpha P_{\mathrm{beam}} \left<\cos{\theta^{*}}\right>\,}
  \frac
    {\,\sqrt{N^{\uparrow}(\cos{\theta^{*}})N^{\downarrow}(-\cos{\theta^{*}})\,}
  -\sqrt{N^{\downarrow}(\cos{\theta^{*}})N^{\uparrow}(-\cos{\theta^{*}})\,}\, }
  {\,\sqrt{N^{\uparrow}(\cos{\theta^{*}})N^{\downarrow}(-\cos{\theta^{*}})\,}
  +\sqrt{N^{\downarrow}(\cos{\theta^{*}})N^{\uparrow}(-\cos{\theta^{*}})\,}\, },
  \label{eq:dttAsy}
\end{equation}
\end{widetext}
where $\alpha_{\Lambda}=0.642\pm0.013$~\cite{Tanabashi:2018pdg}, $\alpha_{\bar{\Lambda}} = -\alpha_{\Lambda}$, $P_\mathrm{beam}$ is the beam polarization and $\left\langle\cos{\theta^{*}}\right\rangle$ denotes the average value in the $\cos{\theta^{*}}$ interval.
$N^{\uparrow(\downarrow)}$ is the $\Lambda$ yield in the corresponding $\cos\theta^*$ bin when the proton beam is upward (downward) polarized.
The relative luminosity between $N^{\uparrow}$ and $N^{\downarrow}$ is cancelled in this cross-ratio asymmetry. 
The acceptance is also cancelled as the acceptance in a small $\cos\theta^*$ interval is expected to remain the same when flipping the beam polarization\,\cite{09DLL:2009xg}, and the effect from limited $\cos\theta^*$ bin width is negligible. 
At RHIC both proton beams are polarized, and the single spin hyperon yield $N^{\uparrow(\downarrow)}$ was obtained by summing over the spin states of the other beam, as the beam fill patterns are nearly balanced with opposite polarizations.

The yields $N^{\uparrow}$ and $N^{\downarrow}$ were first determined in each $\cos{\theta^{*}}$ interval from the observed $\Lambda$ or $\bar{\Lambda}$ candidate yields in the chosen mass interval, and the raw spin transfer, $D^{\mathrm{raw}}_{\mathrm{TT}}$, was extracted using Eq.\,\ref{eq:dttAsy} in each $\cos{\theta^{*}}$ bin.
Then the $D_{\mathrm{TT}}^{\mathrm{raw}}$ values were averaged over the entire $\cos{\theta^{*}}$ range in each hyperon $p_{\mathrm{T}}$ bin.
Since single spin yields can be obtained with either beam polarized, we have two independent measurements for the same physics $\eta$ range, but with different detector coverage, by treating one beam as polarized and summing over the polarization states of the other beam.
After confirming their consistency, the final $D_{\mathrm{TT}}^{\mathrm{raw}}$ results are determined from the weighted mean.
\begin{figure}[htbp]
  \centering
  \includegraphics[width=0.48\textwidth]{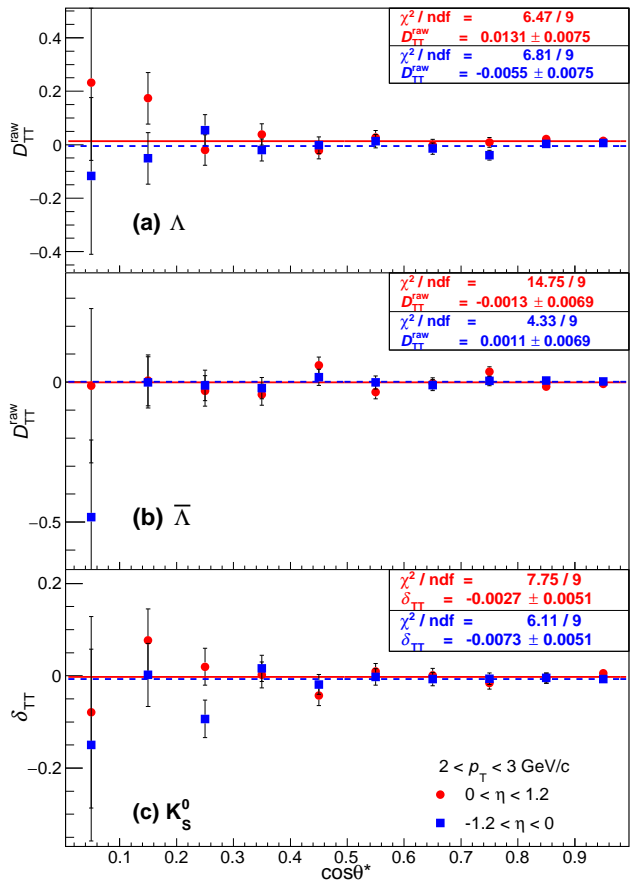}
  \caption{The spin transfer $D_{\mathrm{TT}}^{\mathrm{raw}}$ versus $\cos\theta^*$ for (a) $\Lambda$ and (b) $\bar{\Lambda}$ hyperons, and (c) the spin asymmetry $\delta_{\mathrm{TT}}$ for the control sample of $K_S^0$ mesons versus $\cos{\theta^{*}}$ in the $p_{\mathrm{T}}$ bin of (2,3)\,$\mathrm{GeV}/c$. The red circles show the results for positive pseudo-rapidity $\eta$ with respect to the polarized beam and the blue squares show the results for negative $\eta$. Only statistical uncertainties are shown.}
  \label{fig:dttCos}
\end{figure}

Figure\,\ref{fig:dttCos}(a) shows the beam combined $D_{\mathrm{TT}}^{\mathrm{raw}}$ versus $\cos{\theta^{*}}$ for $\Lambda$ hyperons in the positive and negative $\eta$ regions, respectively, with $2<p_{\mathrm{T}}<3\,\mathrm{GeV}/c$ provided as an example. 
The results for $\bar{\Lambda}$ hyperon are shown in Fig.\,\ref{fig:dttCos}(b).
Positive $\eta$ is defined along the direction of the incident polarized beam. 
The extracted $D_{\mathrm{TT}}^{\mathrm{raw}}$ is constant with $\cos{\theta^{*}}$ as expected and confirmed by the quality of the fit. 
A null-measurement was performed with the spin transfer for the spinless $K_{S}^{0}$, $\delta_\mathrm{TT}$, as a cross check, which has a similar event topology. 
The $\delta_\mathrm{TT}$ obtained with an artificial weak decay parameter $\alpha_{K_S^0}=1$ are shown in Fig.\,\ref{fig:dttCos}(c). 
As expected, the results were found to be consistent with no spin transfer.

$D_{\mathrm{TT}}^{\mathrm{raw}}$ values and their statistical uncertainties were then corrected for residual background dilution according to
\begin{equation}
  D_{\mathrm{TT}}
  =
  \frac
  {\,D_{\mathrm{TT}}^{\mathrm{raw}}-rD_{\mathrm{TT}}^{\mathrm{bkg}}\,}
  {1-r},
  \label{eq:bkgCorrection}
\end{equation}
\begin{equation}
  \delta D_{\mathrm{TT}} = \frac{\sqrt{(\delta D^{\mathrm{raw}}_{\mathrm{TT}})^{2}+(r\delta D^{\mathrm{bkg}}_{\mathrm{TT}})^{2}}}{1-r},
  \label{eq:bkgCorrect_error}
\end{equation}
where $r$ is the background fraction in each $p_{\mathrm{T}}$ bin estimated by the side-band method\,\cite{Adam:2018kzl}, and $D_{\mathrm{TT}}^{\mathrm{bkg}}$ was obtained from the left and right side-band mass intervals as shown in last two rows of Table\,\ref{tab:NLA_J1} using the same procedure as $D_{\mathrm{TT}}^{\mathrm{raw}}$ following Eq.\,\ref{eq:dttAsy} and was consistent with zero within the statistical uncertainty.

The systematic uncertainties of $D_\mathrm{TT}$ include contributions from the decay parameter $\alpha$, from the measurement of the beam polarizations as well as uncertainty caused by the residual background fraction, event pile-up effects and trigger bias due to trigger conditions. 
The total systematic uncertainties in $D_{\mathrm{TT}}$ range from 0.0003 to 0.009 in different $p_\mathrm{T}$ bins, while the corresponding statistical uncertainties vary from 0.006 to 0.040.
The above contributions are considered to be independent and their sizes have been estimated as described below.  
The trigger bias is the dominant contribution to the systematic uncertainty. 

The uncertainty in $\alpha_{\Lambda}$\,=\,$-\alpha_{\bar{\Lambda}}$\,=\,$0.642\pm0.013$ corresponds to a $2\%$ scale uncertainty in $D_{\mathrm{TT}}$.
The uncertainty in the RHIC beam polarization measurements contribute an additional $3.4\%$ scale uncertainty in $D_{\mathrm{TT}}$.
The pile-up effect due to possible overlapping events recorded in the TPC was studied by examining the hyperon yield per event versus the STAR collision rate. 
A comparison between the yields per collision event found by fitting with a constant and a linear extrapolation to very low collision rates was used to estimate the pile up contribution for different spin states, and the corresponding uncertainty to $D_{\mathrm{TT}}$ was found to be small ($<0.005$).
The residual background fractions were estimated by the side-band method and $D_{\mathrm{TT}}$ was corrected accordingly.   
The corresponding uncertainty was quantified from the difference in the results when the background fractions were estimated instead by fitting the mass spectra.  This part is found to be also very small ($< 0.003$).
The data samples used in this analysis were recorded with jet-patch trigger conditions, which help to reach high transverse momenta of hyperons. 
However, these trigger conditions may bias the $D_{\mathrm{TT}}$ measurements by changing the natural composition of hyperon events with respect to the distributions of the hyperon momentum fraction in its associated jet, the hard production subprocesses, the flavor of the associated jet, the decay contribution, {\itshape etc}.
The uncertainties caused by the above mentioned jet trigger conditions were studied with a Monte Carlo simulation of events generated using PYTHIA 6.4\,\cite{PYTHIA:2006za} with the Perugia 2012 tune\,\cite{PerugiaTunes:2010ak} and the STAR detector response package based on GEANT 3\,\cite{geant3}.
The uncertainties from trigger bias were then estimated by comparing the $D_{\mathrm{TT}}$ values obtained from a theoretical model\,\cite{polarizationLA:2005ru} before and after applying the trigger conditions. This yields uncertainties up to 0.008 with increasing $p_\mathrm{T}$.
\begin{figure}[htbp]
  \centering
  \includegraphics[width=0.49\textwidth]{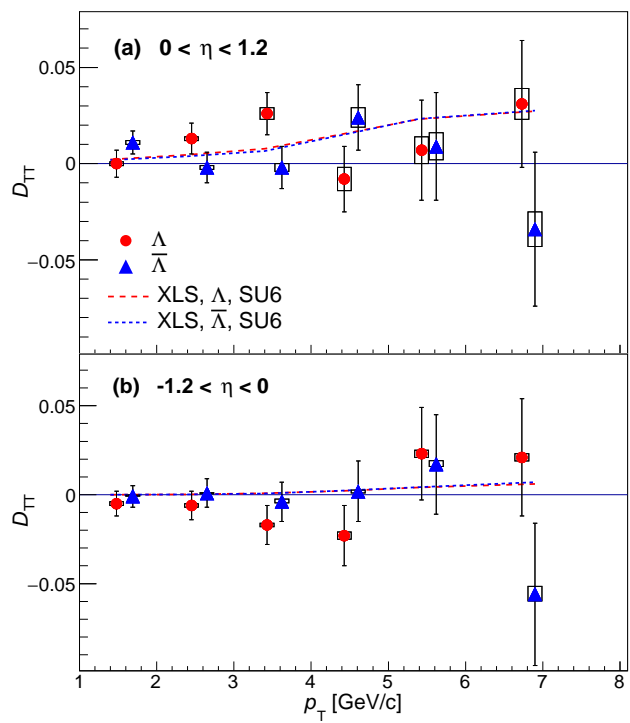}
  \caption{
  The spin transfer $D_\mathrm{TT}$ for $\Lambda$ and $\bar{\Lambda}$ versus $p_\mathrm{T}$ in polarized proton-proton collisions at $\sqrt{s}=200\,\mathrm{GeV}$ at STAR, in comparison with model predictions\,\cite{Xu:2004es,polarizationLA:2005ru} for (a) positive $\eta$ and (b) negative $\eta$.
  The vertical bars and hollow rectangles indicate the sizes of the statistical and systematic uncertainties, respectively. The $\bar{\Lambda}$ results have been offset to slightly larger $p_{\mathrm{T}}$ values for clarity.
  }
  \label{fig:dttPt}
\end{figure}

The STAR results for $D_\mathrm{TT}$ versus hyperon $p_\mathrm{T}$ are shown in Fig.\,\ref{fig:dttPt} for $\Lambda$ and $\bar{\Lambda}$ at both positive and negative $\eta$ regions relative to the polarized beam in proton-proton collisions at $\sqrt{s}=200\,\mathrm{GeV}$. 
About 60\% of $\Lambda$ or $\bar{\Lambda}$ are not primary particles, but stem from decay of heavier hyperons. 
No corrections have been applied for decay contributions from heavier baryonic states. 
Several of the models do take into account the decay contribution\,\cite{deFlorian:1998am,Xu:2004es,polarizationLA:2005ru}.
The statistical and systematic uncertainties are shown with vertical bars and open boxes, respectively.
The systematic uncertainties in the positive $\eta$ range are larger than those in the negative $\eta$ range, owing to the larger trigger bias in the positive $\eta$ range.
These results provide the first measurements on transverse spin transfer of hyperons at a high energy of $\sqrt{s}=200$ GeV.
The $D_{\mathrm{TT}}$ results for $\Lambda$ and $\bar{\Lambda}$ are consistent with zero within uncertainties. 
The data cover $p_\mathrm{T}$ up to $7\,\mathrm{GeV}/c$, where $D_{\mathrm{TT}} = 0.031 \pm 0.033(\mathrm{stat.}) \pm 0.008(\mathrm{sys.})$ for $\Lambda$ and $D_{\mathrm{TT}} = -0.034 \pm 0.040(\mathrm{stat.}) \pm 0.009(\mathrm{sys.})$ for $\bar{\Lambda}$ at $\left<\eta\right> = 0.5$ and $\left<p_{\mathrm{T}}\right> = 6.7\,\mathrm{GeV}/c$. 
Strange quarks and anti-strange quarks are expected to carry a significant part of the spins of $\Lambda$ and $\bar{\Lambda}$; therefore the measurements of transverse spin transfer to them can provide insights into the transversity distribution of strange quarks.
Knowledge of transversity of valence quarks has been learned mostly from DIS experiments.
But the transversity distribution of strange quarks is not yet constrained experimentally\,\cite{Anselmino:2013vqa,Kang:2015msa,Radici:2018iag}. 

A few model predictions exist of hyperon transverse spin transfer in hadron-hadron collisions, based on different assumptions on the transversity distribution and transversely polarized fragmentation functions\,\cite{deFlorian:1998am,Xu:2004es,polarizationLA:2005ru}.
In Fig.\,\ref{fig:dttPt}, the $D_{\mathrm{TT}}$ data are compared with a model estimation calculated at $\left<\eta\right>=\pm 0.5$, which is available for RHIC energy, with simple assumptions of the transversity (using the DSSV helicity distribution as input) and the SU6 picture of fragmentation functions\,\cite{Xu:2004es,polarizationLA:2005ru}.
The data are in general consistent with the model predictions.
The $D_{\mathrm{TT}}$ of $\Lambda$ and $\bar{\Lambda}$ are not in disagreement with each other, with a $\chi^{2}/$ndf of 9.5/6. 
A possible difference between them could come, for example, from different fragmentation contributions from the non-strange quarks.

In summary, we report the first measurement of the transverse spin transfer, $D_{\mathrm{TT}}$, to $\Lambda$ and $\bar{\Lambda}$ in transversely polarized proton-proton collisions at $\sqrt s$\,=\,200\,GeV at RHIC. 
The data correspond to an integrated luminosity of 18 pb$^{-1}$ taken by the STAR experiment in 2012, and cover mid-rapidity, $\left|\eta\right| < 1.2$, and $p_{\mathrm{T}}$ up to $8\,\mathrm{GeV/}c$. 
The $D_{\mathrm{TT}}$ value and precision in the highest $p_{\mathrm{T}}$ bin, where the effects are expected to be largest, are found to be $D_{\mathrm{TT}} = 0.031 \pm 0.033(\mathrm{stat.}) \pm 0.008(\mathrm{sys.})$ for $\Lambda$ and $D_{\mathrm{TT}} = -0.034 \pm 0.040(\mathrm{stat.}) \pm 0.009(\mathrm{sys.})$ for $\bar{\Lambda}$ at $\left<\eta\right> = 0.5$ and $\left<p_{\mathrm{T}}\right> = 6.7\,\mathrm{GeV}/c$.  
The results for $D_\mathrm{TT}$ are found to be consistent with zero for $\Lambda$ and $\bar{\Lambda}$ within 
uncertainties, and are also consistent with model predictions.

We thank the RHIC Operations Group and RCF at BNL, the NERSC Center at LBNL, and the Open Science Grid consortium for providing resources and support.  This work was supported in part by the Office of Nuclear Physics within the U.S. DOE Office of Science, the U.S. National Science Foundation, the Ministry of Education and Science of the Russian Federation,
Natural Science Foundation of China, Chinese Academy of Science, the Ministry of Science and Technology of China, the Major State Basic Research Development Program in China (No.~2014CB845406) and the Chinese Ministry of Education, the National Research Foundation of Korea, Czech Science Foundation and Ministry of Education, Youth and Sports of the Czech Republic, Department of Atomic Energy and Department of Science and Technology of the Government of India, the National Science Centre of Poland, the Ministry  of Science, Education and Sports of the Republic of Croatia, RosAtom of Russia and German Bundesministerium fur Bildung, Wissenschaft, Forschung and Technologie (BMBF) and the Helmholtz Association.

\bibliographystyle{apsrev4-1}
\bibliography{pp200_2012_SpinTransfer}

\end{document}